\documentclass[12pt]{article}
\begin{document}
\title{Reminiscences of Julian Schwinger:\\Late Harvard, Early
UCLA Years (1968-1981)}

\author{K. A. Milton}

\date{\today}
\maketitle

I came to Harvard as a graduate student in 1967, and the first year I was just 
taking courses, not very good ones at that.  It was only in the Fall of 1968 
that I sat in on a course from Julian, Physics 251, Quantum Mechanics, although
 I had taken that course the previous year from Frank Pipkin.  I was, of 
course, blown away by Julian's elegance and insight.  Although I had thought 
momentarily of asking Curt Callan to be my research advisor, I quickly realized
 that he was but an Assistant Professor and would therefore not be granted 
tenure by Harvard, and so would not be around long enough for me to finish.  
Sydney Coleman had no ideas he was willing to share with students, and Shelly 
Glashow was all over the place (although he had just guided Tim Boyer to a 
heroic calculation of the Casimir effect for a sphere, about which more later).
  So there was no choice but for me to work with Julian.  You see I had no idea
 who was right for me.  I had even never heard of Schwinger before I came to 
Harvard, unlike the icon of Feynman.

Julian was very eager to take on students, and I immediately joined his group 
of twelve disciples.  He had taken the advice he had given in his Nobel Lecture
 in 1965 and invented a new approach to field theory in which the infinities 
were apparently removed, Source Theory.  At that point, 1968, he had two papers
 out on the subject \cite{Schwinger:1966zz}, 
including one on quantum electrodynamics \cite{Schwinger:1967rg}, 
as well as a few
 short notes on effective Lagrangians for chiral symmetry
\cite{Schwinger:1967tc}, so as usual, the 
students had to learn most of the subject from the master's knee, so to 
speak.  In my case I had missed the source theory course in 1967--68, but I got
the notes from Wu-yang Tsai, and went through them in detail.  In 1968--69 I 
was still taking courses, but 75\% of my time was devoted to sourcery.

In those days, Schwinger lectured three days a week, nominally 12:00--1:00, 
but he 
invariably was late, arriving in his Iso Rivolta about 12:15, and the lectures 
ran later and later, approaching 2pm.  I remember one occasion when Julian was 
teaching Quantum Mechanics, a course in which half the audience were 
undergraduates, whose Houses stopped serving lunch at 2.  One day he asked, 
rhetorically, ``If I could just have a few minutes more?'' and was met with a 
Harvard hiss.  He stormed out, only to encounter a locked door, so he had to 
slink out the side, but never again (at least that year) lectured beyond 1:30.

His dozen graduate students could only meet him on Wednesday afternoon, after 
he returned from lunch.  One year Roman Jackiw was visiting Harvard, so they 
often had lunch together, and sometimes they never returned.  Julian's 
secretary would put out a list for students to sign up to see him at 9am, and 
we would stagger in as close as possible to that time (since we had stayed up 
all night trying to prepare for our audience with the great man) because if 
you were near the top of the list, you would likely get in to see him before he
 left at 6, but if you were number 10, you would probably fail to do so.  Once 
admitted to his sanctum, Julian gave you his complete attention, of course, 
ignoring the telephone, and allowing you to explain what you had been doing the
 last week or two, and getting up to where you were stuck. Invariably, he would
 make an insightful suggestion of how to proceed.  It might not always work, 
but it would take a week at least to follow the suggestion 
through.  There was never any 
feeling of time constraint in the meetings, again proving the necessity for 
being at the top of the sign-up list.

Early on, Julian suggested I look at spectral forms for scattering, what were 
conventionally called dispersion relations, but derived by looking at causal 
processes and then performing ``space-time extrapolation.''  It was a 
very effective method of getting results.  My first real project was rederiving
 the Lamb shift using these methods, a nice pedagogical project.  I was to 
present that at my oral examination, probably in Spring 1969.  I had just 
barely got launched, talking about the basic concepts of source theory I would 
use in the calculation when Paul Martin objected, and before I could say 
anything, Julian responded, and then went to the board for a 30 minute 
exposition.  I think I may have been allowed to say a few words about my 
conclusions, but then I was ushered out for the committee's deliberation.  
I must have passed, because Julian gave me a copy of his just-published 
Brandeis lectures on source theory \cite{brandeis}.

The other topic he suggested for my thesis was looking at the quantum coupling 
of electrons and photons to gravity.  This was about the time of the discovery 
of Callan, Coleman, and Jackiw of the ``new, improved''  stress 
tensor for scalar fields, the conformal stress tensor \cite{Callan:1970ze}.  
I again investigated this using spectral forms.  At first, I had not defined 
the basis tensors quite correctly, so I asked a stupid question of Sydney 
Coleman at the Erice Summer School in 1970, because I didn't initially see the 
advantage of the conformal tensor.  But fortunately I discovered the error 
while I was completing my thesis, and my published results were an early 
manifestation of the trace anomaly \cite{Milton:1972cc}.

In the Spring of 1970, Julian and Clarice took a sabbatical in Japan.  To keep 
his group together, Richard Ivanetich, his student who had become another 
non-permanent Assistant Professor at Harvard, gave a course on source theory, 
and so we kept ourselves going.  But when Julian returned in the Fall, 
having completed his first source-theory book \cite{Schwinger:1970xc},
he met his assembled students with a bombshell:  He had finally accepted David 
Saxon's entreaties and had decided to move to UCLA, much to the dismay for his 
Boston-born wife Clarice.  We students were all greatly disheartened.  But 
then, when I had my private meeting with him, Julian explained that he had 
arranged for me to come along as a postdoc, together with Lester DeRaad, Jr., 
and Wu-yang Tsai.  The latter were really more than ready to defend, but I was 
stunned, because I had expected to have another year to finish up. 
I was not ready to leave civilized Cambridge for the wilderness of California.
 I worked 
hard that Fall, and moved out to California in February 1971, where I finished 
writing up my thesis.  Julian and Clarice had arrived a week earlier, to be
greeted by the great San Fernando earthquake.

Why did Julian decide to leave Harvard?  The fact that the reception of source
theory by his colleagues and former students at Harvard was rather frosty 
was a contributor, but Julian always stated positive reasons: He wanted to move
to a gentler climate, where he could play tennis every day, and swim, as well
as ski occasionally.  Formerly famously overweight, he had become obsessed with
becoming more fit after his imperfect idol Wolfgang Pauli succumbed to 
pancreatic cancer in 1958. The Schwingers bought a house in Bel Air, with a
beautiful view of the city of Los Angeles, of course with a pool.  Clarice's
mother, Sadie, had her own apartment in their house.  Saxon and Schwinger
both presumed that students would continue to flock to him on the West Coast:
he had had some 70 students at Harvard, and was to spend the last 23 years of
his life at UCLA, but the influx of students was not to be.  The students there were clearly not
so talented as the Harvard coterie, and by 1971 Julian's work had fallen out
of fashion.  At UCLA, Julian had no more than five students, including
Walter Wilcox, who became my first postdoc at Oklahoma State,
now at Baylor, and Luis Fernando
Urrutia, now professor at UNAM in Mexico City.

My thesis defense was even less trying than my oral exam.  By this point, 
Julian and I had decamped to UCLA, and Julian took the local
committee and me to a 
French restaurant.  (Harvard had  very flexible rules about the composition
of doctoral committees.) 
I successfully answered Julian's only question, 
``where were you born?,'' and from that point I only had to deal with the 
hazards of long-distance transmittal of my dissertation to Harvard, navigating 
the new technologies of the Xerox machine (Harvard had a secret, unpublished 
rule about that), and the lack of FedEx in those days!

I officially became an Assistant Research Physicist/Assistant Visiting 
Professor at UCLA in Fall 1971.  In those days, almost all postdocs at UCLA 
taught two courses per year.  (My future father-in-law, Alfredo Ba\~n±os, Jr., 
who became Vice-Chair, once published a chart showing the monotonic decrease 
of the student faculty ratio with faculty rank, with the postdocs having the
heaviest teaching load.)  But my higher status did mean that I had much more 
intimate contact with Julian; whereas as a student I only saw him as a 
brilliant lecturer and bi-weekly mentor, at UCLA we had weekly lunches at 
various venues, and we were occasionally invited to the Schwingers' home.  
These informal meetings were wonderful, full of discussions of physics and 
its culture. At dinner, 
Julian was a gracious host, drawing out quiet individuals to join in the 
conversation.  He was never one to flaunt his deep erudition.  I remember one 
dinner where he sat next to my wife's nephew who was a sulky teenager, and 
Julian got him into the conversation by asking him about Thelonious  Monk, one 
of the boy's heroes.

Not only did more social contact occur after the phase transition into being a 
postdoc, but also collaboration ensued.  Julian very seldom worked on projects 
his students were involved in, but working with his ``assistants'' was another 
matter.  So in 1974 when the discovery of the $J/\psi$ particle was announced, 
Julian immediately came up with an explanation, in terms of a 
previously hidden sector \cite{Schwinger:1974pr},
and suggested this might have something to do with dyons 
\cite{Schwinger:1975km}, particles having both
 magnetic and electric charge.  Although these explanations ultimately proved 
unsatisfactory, and fell to the idea that the $J/\psi$ was a bound state of 
charmed and anti-charmed quarks,\footnote{Julian detested quarks, partly
because Gell-Mann had invented them.  He provided a critique of their
naming in his earlier Science article 
proposing a magnetic model of matter \cite{Schwinger:1969ib}.} 
 Julian did invite his postdocs into a convincing explanation of the decay 
$\psi(3.7)$ into the $J/\psi$ state \cite{Schwinger:1975ps}. 
(This work resulted in an ugly break in relations with former
student Asim Yildiz.)  And after giving 
a heuristic derivation of precession tests in general relativity
\cite{ajp1,ajp2}, he invited me
 to follow on with an analysis of the Lense-Thirring effect \cite{ajp3}.  
After the source theory book he wrote while he was on sabbatical in Tokyo in 
1970 \cite{Schwinger:1970xc}, Julian invited us three postdocs 
(and likely Jack Ng, who had come from Harvard to UCLA to finish his Ph.D. with
 Julian) to proofread the second volume.  It was inadvertence, not an 
intentional slight, that led him to neglect to thank us in the preface to the 
resulting volume, which came out in 1973 \cite{Schwinger:1973rv}.  We jointly 
wrote a very pretty paper on nonrelativistic dyon-dyon scattering 
\cite{Schwinger:1976fr}, which included as co-author D. C. Clark, one of 
Julian's few UCLA students.

Personally, a stellar point of my life was my introduction to Margarita 
Ba\~nos by Clarice.  Julian and Margarita's father Alfredo Ba\~nos had been 
colleagues at the Radiation Laboratory at MIT during World War II, working 
together on the theory of radar.  This work proved crucial for Julian's later 
discovery of renormalization in quantum electrodynamics, which was the basis of
 his Nobel Prize \cite{Schwinger:1982bm}.  
Alfredo had moved to UCLA soon after the war, as had David Saxon, who had 
written up Julian's lectures on microwave theory given at the end of the war, 
which eventually, in edited form, came to life in {\it Electromagnetic 
Radiation} \cite{Milton:2006ia}.  In 1972 Clarice had arranged a dinner party 
for my postdoctoral colleague Lester DeRaad and Margarita, which did not go 
well, Lester particularly objecting to the blind date.  (Margarita and Lester 
are now on good terms.)  Some six weeks later, Clarice tried again with me, but
 this time just suggesting I call Margarita for a date.  Don the Beachcomber, 
and a {\it voyage surprise\/} through the Santa Monica mountains later (with 
an interruption for Margarita dancing in New York), Margarita and I became 
a couple, married in 1978.

Schwinger's last substantial work in high-energy physics concerned an analysis 
of the scaling properties observed at Stanford in deep-inelastic scattering of 
electrons and neutrinos on nucleons \cite{dis1,dis2}.  
Rather than interpreting these in terms 
of Feynman's partons, which later got conflated with quarks, he adopted a more 
phenomenological viewpoint, describing them in terms of double spectral forms, 
related to the Deser-Gilbert-Sudarshan representation 
\cite{Schwinger:1975ti,Schwinger:1976ix,Schwinger:1976dw,Schwinger:1977re,
Schwinger:1977rfa}.  
We postdocs joined in \cite{Tsai:1975tj,DeRaad:1975jj}, but
eventually, toward the end of our extended tenure at UCLA, we  discovered
 that although the structure of the spectral forms was 
valid, in general the spectral region was not confined to positive mass 
distributions, which rendered conclusions suspect \cite{Ivanetich:1978pt}.

Somewhat earlier,
he had returned to the subject he had mastered during the war, synchrotron 
radiation, adopting a quantum viewpoint \cite{Schwinger:1973kp}, and this work 
led to collaborative papers with Tsai and Tom Erber 
\cite{Schwinger:1977ba,Schwinger:1974rq}, who was a frequent visitor to our 
group at UCLA in the mid 1970s, invariably referring to Schwinger as (Big) 
Julie, reminiscent of what  Oppenheimer had done back in the 1940s.

Just before his ``source-theory revolution'' Schwinger had written on the 
subject of magnetic charge 
\cite{Schwinger:1966zza,Schwinger:1966zzb,Schwinger:1966nj}. 
He immediately thereafter put the theory in source theory language
\cite{Schwinger:1968rq},
and followed this with a proposal that matter is composed of dyons
\cite{Schwinger:1969ib}, a name he coined to describe 
particles carrying both electric and magnetic charge, an intriguing
idea in place of quarks.
 The subject of magnetic charge  he revisited 
in 1975 \cite{Schwinger:1975ww}, bemoaning, ``If only the 
Price had been right!'', referring to Buford Price's
discredited claim of discovering magnetic monopoles in Lexan sheets
exposed to cosmic rays \cite{price}.
  (Julian often worked with the TV on.)  The following year, 
with such experiments in mind, we jointly revisited how magnetic charge 
interacted with electric charges, in a monumental paper on dyon-dyon 
scattering \cite{Schwinger:1976fr}.  
Two decades later, this work led to a new experiment with George Kalbfleisch
\cite{Kalbfleisch:2003yt}, 
in turn inspired by Luis Alvarez' searches in lunar samples \cite{Ross:1973it};
our work set the best lower limit on magnetic monopole masses for a decade 
until LHC data supplanted them \cite{pdg,Acharya:2016ukt}.

The reception of the high-energy physics community to Schwinger's source-theory
program was not warm; when not overtly hostile, the ideas were largely ignored.
Julian's reaction was to become ever more iconoclastic.  He developed his own
approach to the renormalization group, based on the photon propagation 
function \cite{Schwinger:1975pnas,Schwinger:1975th}, 
reflecting his rejection of the the whole notion of renormalization,
a concept which he had largely invented.  (A more confrontational paper was 
never published; he increasingly turned to publishing in the Proceedings of
the National Academy, where he could publish without encountering hostile
reviewing.)  Although these ideas were largely rejected by the physics world,
they did later spark some important insights into the running of the strong
coupling constant in QCD \cite{Milton:1996fc}. An even more confrontational
issue developed concerning the decay of the neutral
pion into two photons, $\pi^0\to
\gamma\gamma$.  This had been initially explained by Schwinger back in 1951
in his most famous paper \cite{Schwinger:1951nm}; 
it is a manifestation of the axial-vector or chiral anomaly.  This
subject got rediscovered in the late 1960s; and in particular Adler and
Bardeen proved that the anomaly was not subject to radiative corrections,
but was given exactly by the lowest-order triangle diagram \cite{Adler:1969er}.
In 1972, we postdocs showed that the situation was rather more subtle, in 
that the pion decay process did possess higher-order QED corrections unless
the pseudoscalar coupling is normalized at an infrared sensitive point
\cite{Deraad:1973ee}.  The
formal Adler-Bardeen theorem can, however, be maintained.  Julian subsequently
redid our calculation in his own, inimitable way, and obtained the same
result that we did, a correction by a factor of $1+\alpha/(2\pi)$, where
$\alpha$ is the fine-structure constant.  But instead of seeking an 
accommodation with received wisdom, he chose to fight, and we wrote a joint
paper.  However, before we submitted it, Julian gave a talk at MIT on the 
subject, and was met with utter disdain.  The paper was therefore never
submitted to a journal, but a rather contentious section (subtitled ``A
Confrontation'') of the
third (uncompleted) volume of {\it Particles, Sources, and Fields}, now
included in the repackaged version of the latter part of the second volume
plus the beginnings of the third, contains his
iconoclastic calculation \cite{psfIII}.  This unfortunate lost battle can
be said to mark Schwinger's end of involvement in high-energy physics proper,
and the end of his attempt to complete his source theory program to include
strong interactions.

But Julian certainly did not reject all new ideas.  He became fascinated by
the ideas of supersymmetry, and its local version, supergravity, and in 1977
invited his former distinguished student Stan Deser, who had recently been
one of the co-discoverers of supergravity, to UCLA to give a week of
private lectures to his group, including Bob Finkelstein.  Julian regretted
he had not thought of the idea of fermion-boson unification, particularly
since he had long before set up the key ingredients, such as the multispinor 
formulation and Grassmann variables.  After this command performance 
Julian wrote his own version of supersymmetry \cite{Schwinger:1978ra};
Bob and I followed with a rederivation of supergravity \cite{Milton:1978qs}.
But this reconstructive work had negligible impact.
 
Most important for my later career, Julian learned about the Casimir effect 
from Seth Putterman, and immediately set about seeing how he could derive it 
without using the offensive idea of zero-point energy. After a short solo note
\cite{casimir}, this led to two very substantial joint papers, one devoted to 
rederiving the Casimir effect between parallel dielectric slabs, the so-called 
Lifshitz theory \cite{dielectrics},
 and the second to rederiving the result of Tim Boyer, as a student of 
Glashow at Harvard, that the Casimir self-stress on a perfectly conducting 
spherical shell of zero thickness is repulsive, not attractive as Casimir and 
everyone else had supposed \cite{Milton:1978sf}.  This is a subject I have 
never left.

During all this time at UCLA Julian taught brilliant courses.  Of course, 
quantum mechanics, now for undergraduates, based on his Measurement Algebra 
approach.  Berge Englert eventually turned this into a book \cite{englert}.  
And, for the first time since the war, he taught the graduate Classical 
Electrodynamics course.  After we three postdocs had sat in on that for a 
while, we asked if we could write up the notes into a book.  A draft was sent 
to W.H. Freeman, who accepted the proposal; this caused Julian to pay 
attention.  He looked at our manuscript, decided he could greatly improve upon 
it, and spent the next decade doing so, teaching the subject several times,
 before abandoning the project before reaching radiation theory.  So, after 
his death in 1994, I undertook to turn the old typescript and the 
multitudinous revisions into modern \LaTeX\ form; the book was published in 
1999 \cite{ce}, and has had modest success, but has hardly dented sales of 
Jackson \cite{jackson}.

Julian also wanted to communicate the excitement of science to a wider 
audience.  His most notable efforts in this regard was in collaboration with
the BBC, where, with George Abell, a UCLA astronomer, he designed a course
on relativity, both special and general, for the Open University, entitled
``Understanding Space and Time.''  Julian was very excited about this project,
working on it from 1976--79, which resulted in a number of TV programs
aired on BBC2, and occasionally in the United States.  In Los Angeles they
were aired in the early morning by KCET, not surprising since the release
roughly coincided with the extremely successful KCET {\it Cosmos\/} series (1980),
also co-produced by the BBC,
hosted by the extremely engaging Carl Sagan.  In contrast, Julian's TV style 
appeared rather wooden.  A lasting testament to this endeavor was the book
{\it Einstein's Legacy} \cite{EL}, aimed at a popular audience.

Julian, although he put in long hours at home working (never allowing himself
to be interrupted by a telephone call), had outside interests.  We have already
mentioned his pleasure in frequent skiing trips, and he had weekly tennis
matches (which he had learned from his student Asim Yildiz at Harvard) with
Lester. Privately, he enjoyed playing the piano, but never when anyone was
around.  (``Anything worth doing, is worth doing badly,'' except physics, of
course.) And in 1975 he became the second-largest owner in a vineyard in 
Northern California, the V. Sattui Winery, which has been remarkably
successful since its relaunch in 1976.

1978 marked Schwinger's 60th birthday.  The UCLA Physics Department organized a
 Fest in Julian's honor.  I ended up being defacto chief organizer of that 
meeting, which, not surprisingly, included many luminous names. It resulted
in a rather nice Festschrift volume \cite{fest}, but the transcript
of the wonderful talk given by Feynman at the banquet was not published
until the Festshrift for Julian's 70th birthday came out \cite{fest2}.
 Julian was quite grumpy about the whole affair, because he saw it as a sort of
retirement celebration.  A few years later, at a meeting in Atlanta, where he 
received the Monie Ferst award from Sigma Xi, he publicly apologized to 
his former students, me,
Ken Johnson of MIT,  and Margie Kivelson of UCLA, for being ungracious.

It is often asked whether Feynman and Schwinger got along.  Certainly, they
always behaved cordially at conferences, both recalling how in the early
days of QED only by comparing each other's completely orthogonal calculations
and seeing that they yielded precisely the same predictions were they 
themselves convinced that their tentative procedures were sound.  But it is 
true that
for nearly two decades the two shared Nobel laureates lived in the same
metropolis and never socialized privately.\footnote{Berge Englert has told
me that the Schwingers once invited Dick and his wife to dinner, which Dick
greatly enjoyed until a second couple appeared.  This  spoiled the evening
for the Feynmans, who had expected they were to be the only guests.}
Feynman made several overtures,
suggesting that they meet at a restaurant somewhere 
between Pasadena and Bel Air,
but Julian was stand-offish, and such an encounter never happened, I think to
both men's deep regret.  This failure to connect was another mark of Julian's
extreme privacy.

Eventually the group of ``sourcerer's apprentices'' at UCLA broke up.  Tsai 
left first to take up a faculty job a Coral Gables, returning to Southern 
California after a year to work at JPL.  DeRaad left for a career in industry, 
first working for R\&D Associates, and attempted with Tsai to found a 
defense-based firm in the 1980s, with Julian on the board of directors.  
I could have stayed on at UCLA  indefinitely, but with little chance of a 
faculty appointment, I followed my wife Margarita to Ohio State, and then, 
after two years, landed a faculty job in Oklahoma State.  Three years later 
Margarita got an Assistant Professorship at the University of Oklahoma, and I 
was able to secure a Professorship there in 1986, where I've been based ever 
since.

I met Julian a few times after I officially resigned from UCLA in 1981.  I 
invited him to give a public talk at Oklahoma State in 1984, and I tried to do 
the same when I moved to Norman.  Of course, I attended his 70th Fest in 1988
\cite{fest2}, which was dedicated to the memory of Dick Feynman, who had
just succumbed to cancer. 
 In 1989 came the debacle of cold fusion, and Julian, being ever the 
iconoclast, fell into it with an explanation, which was convincing to very few.
Berge Englert helped him publish one of his papers on this, which was printed 
alongside a publisher's disclaimer \cite{zpd}.  Eventually he came to realize 
his ideas, and cold fusion itself, could not be right, and he turned to a 
subject that undoubtedly had experimental support, sonoluminesence, again at 
the urging of Seth Putterman.  He decided this must have something to do with 
the dynamical Casimir effect, and published several short notes in the PNAS
\cite{sono}. 
 Unfortunately, he forgot that I had written, while still at UCLA,
 a paper on the Casimir energy of a
 dielectric ball \cite{Milton:1979yx}, which was rather the inverse of the 
problem he was considering, a bubble in water, so he developed to a certain 
extent his own treatment.  When I last met him at the annual Christmas party 
given at the Ba\~nos' home in 1993, he discussed his latest efforts in this 
direction, and indirectly suggested I join in.  Tragically, this was not to be,
 because in the following February he was diagnosed with pancreatic cancer.  
After his death I continued to pursue the subject, but eventually, with my 
colleagues Jack Ng and Iver Brevik, demonstrated to our satisfaction, at least,
 that the Casimir effect, dynamical or not, could not be relevant to the 
copious light produced repeatedly during bubble collapse \cite{Milton:1997ky,
Brevik:1998zs}.

So how do we assess the legacy of Julian Schwinger?  He was, of course, a giant
of 20th century physics, who completely dominated theoretical physics for a 
decade in the 1950s. His influence on physics in the 21st century is pervasive,
 even if largely unrecognized by many, 
particularly in the younger generation.  
The fact he had so many brilliant and influential students guarantees the 
impact of his school on future generations. The techniques he invented, from 
the quantum action principle, effective Lagrangians, the Schwinger-Keldysh 
method, commutator anomalies, proper time methods, and ``Feynman'' parameters,
 to name a few, underlie much of modern theoretical physics, and are often used
in ignorance of their inventor.   Of course, he made mistakes and took wrong 
directions, but as Einstein said, ``Anyone who has never made a mistake has 
never tried anything new.''  And to every problem to which Julian turned his 
attention, he brought new insight and new techniques, often sparking
a whole new field of endeavor.  He remains a beacon, 
guiding us into the future.

For more about  Julian Schwinger's life and work see \cite{bio}, with 
some updates in \cite{update}.

{\bf Acknowledgments:}
I thank Berge Englert for asking me to write this reminiscence, and for his
close reading of the manuscript.
I dedicate this note to Julian's memory.

\end{document}